# Comparison of Tumor and Normal Cells Protein-Protein Interaction Network Parameters

HosseinAli Rahmani Dashti[1], Ali Reza Khanteymoori[2] and Mohammad Olyaee[2, *]

[1] *Computer Engineering Department, University of Zanjan, Zanjan, Iran*
srahmani@znu.ac.ir

[2] *Computer Engineering Department, University of Zanjan, Zanjan, Iran*
khanteymoori@znu.ac.ir, mh.olyaee@znu.ac.ir

**Abstract.** In this paper, we compared cancerous and normal cell according to their protein-protein interaction network. Cancer is one of the complicated diseases and experimental investigations have been showed that protein interactions have an important role in the growth of cancer. We calculated some graph related parameters such as Number of Vertices, Number of Edges, Closeness, Graph Diameter, Graph Radius, Index of Aggregation, Connectivity, Number of Edges divided by the Number of Vertices, Degree, Cluster Coefficient, Subgraph Centrality, and Betweenness. Furthermore, the number of motifs and hubs in these networks have been measured.

In this paper bone, breast, colon, kidney and liver benchmark datasets have been used for experiments. The experimental results show that Graph Degree Mean, Subgraph Centrality, Betweenness, and Hubs have higher values in the cancer cells and can be used as a measure to distinguish between normal and cancerous networks. The cancerous tissues of the five studied samples are denser in the interaction networks.

## 1. Introduction

Cancer is one of the complicated diseases which is responsible for almost 13% of all deaths in the world.[1] Several experimental investigations have been showed that genes and protein interactions have an important role in the growth of cancer.[2]
Proteins are necessary macromolecules of cells which are responsible for many biological functions. For some more complicated functions they interact with each other and construct networks which are called Protein-protein interaction (PPI) networks.

A large number of computational methods have been developed to predict PPIs, such as gene neighborhood, gene fusion, gene co-expression, text-mining techniques and etc.[3] PPI data are always represented as a mathematical graph, where each node represents a protein and an interaction between a pair of proteins is indicated by an edge.

Despite many traditional approaches which concentrate on studying genes or proteins, the systematic investigation of differential structure between normal and cancer PPIs may provide a good idea to extract significant biological information for detecting the molecular and cellular mechanisms of cancer and other complex diseases. It is be noted that the gained information can be used for disease diagnoses and treatments.

---

* srahmani@znu.ac.ir

In recent years, many attempts have been made with the purpose of utilizing network analysis in available PPI networks of cancers. For example, the authors in [4] described that the number of hubs in cancerous PPIs can be decreased dramatically in comparison with PPIs in normal tissues. In [5] have studied 22 different graphs which are related for factors for 29 tumors. The authors have considered to the differences of the cancerous networks against random networks. For this purpose, their gene-expression and predicted human protein interactions have been studied. They have found that the outbreak of hub proteins is not increased in the appearance of cancer. In [2] the authors have found that cancer proteins tends to have the higher degree, higher betweenness, shorter shortest-path distance, and weaker clustering coefficient. In [3] the authors investigated that cancer proteins tend to have the higher degree and smaller clustering coefficient against non-cancer proteins. Also cancer proteins have larger betweenness centrality compared to the other proteins.

The present paper aims to call into question the structural differences of normal and cancer protein interaction networks (PINs) by using graph-theoretical concepts and parameters.

It is be noted that some of which have not been examined in previous works.They can be used to predict the properties of the networks. The PINs are related to five tissues i.e. bone, breast, colon, kidney and liver in both normal and cancer conditions.

This paper is organized as follows: The materials and methods that are used for this study and the proposed network construction and analysis methods in both the normal and cancerous tissues are explained briefly in the next section. In the Experimental Results and Discussion section, this study's results and tools has been represented and finally the paper is concluded.

## 2. Materials and Methods

As demonstrated by a series of recent publications [2, 3, 4] to develop a useful analysis method for a biological system, the researchers need to go through five steps: (a) Select a valid dataset; (b) data formulation (c) introduce the proposed method; (d)analysis the results; (e) establish a user-friendly web-server. Below, we are to describe how to deal with these steps one-by-one.

### 2.1. *Materials*

Data set used to evaluate the proposed parameters is the same set used in [6]. This dataset contains normal and cancerous tissues were from several database such as Cancer Cell Map Database, PIPs, and String. There are 609 proteins were found to demonstrate total 8359 possible interactions between them. The expression data were represented in digital expression unit (DEU). In DEU the expressed proteins were assigned values 1 and the unexpressed proteins were assigned values 0. For this purpose, we have constructed PPI data with each pair of protein having 1 in the expression values that is a valid interactions between the proteins.

### 2.2. *Protein-protein interaction networks*

The recent vast amount of experimental PPI data is available which has made graph theory approaches as an important part of computational biology and the knowledge discovery processes. PPIs are commonly represented as a graph which nodes corresponding to proteins and the edges representing the PPIs.

The answer of what is hold essential information and how various proteins act in unison with others to enable the biological processes within the cell is PPIs.

Prediction protein functions are one of the most challenging tasks in the computational biology research. Studying and analyzing of PPIs will provide a valuable insight into the internal mechanism of cells and complex diseases. Moreover, Analyzing PPI networks can provide useful information of the function of individual proteins, protein complexes, and larger subnetworks.

Large-scale and high-throughput are two important techniques which can discover proteins that interact with each other within an organism. Using these techniques are very applicable for PPI data which has been generated are fortunately, many of the public data sets of PPIs are currently available.

Explaining the relationship between structures, functions, and regulations of molecular networks is one of the goals of system biology. It happens by combining theoretical and experimental approaches. Graph theory is an essential part of this process, which enables us to analyze structural properties of PPI networks and specifies other information such as function.

### 2.3. *Network Parameters*

In this work, twelve different network parameters are calculated which include Number of Vertices, Number of Edges, Closeness, Graph Diameter, Graph Radius, Index of Aggregation, Connectivity, Number of Edges divided by the Number of Vertices, Degree, Cluster Coefficient, Subgraph Centrality, and Betweenness. We have calculated the above parameters to analyze their values and differences in different cancerous and normal PPIs.

Therefore, Let $G = (V, E)$ be an undirected graph and $dist(i, j)$ be a shortest path between the node i and j.

#### 2.3.1. *Closeness:*

This parameter shows essential node that can reach quickly with other nodes in the network.[7]

$$C_c(i) = \frac{1}{\sum_{j \in V} dist(i, j)} \quad (1)$$

Sum of $C_c(i)$ indicates graph closeness centrality.[5]

#### 2.3.2. *Graph diameter:*

Diameter is the maximal shortest path distance amongst all the distances calculated between each couple of vertices in graph G.[8]

$$C_G = \max(dist(i,j)) \qquad (2)$$

### 2.3.3. *Graph radius:*

Radius is the minimal shortest path distance amongst all the distances calculated between each couple of vertices in graph G.[8]

$$R_G = \min(dist(i,j)) \qquad (3)$$

### 2.3.4. *Index of aggregation:*

IoA is fraction of the total number of vertices in the Subgraph (A) out of the total number of all given vertices in the graph (B).[5]

$$IoA = \frac{A}{B} \qquad (4)$$

### 2.3.5. *Connectivity:*

Connectivity is derived according to the bellow relation which A is the total number of edges realized in a given graph and B represents the maximum number of possible edges.[5]

$$C = \frac{A}{B} \qquad (5)$$

### 2.3.6. *Subgraph centrality*:

Subgraph Centrality of each node indicates participation of nodes in all subgraphs in the graph.[9]

$$SC_i = \sum_{k=0}^{\infty} \frac{\mu_k(i)}{K!}, \quad \mu_k(i) = \left(A^k\right)_{ii} \qquad (6)$$

### 2.3.7. *Degree*:

The number of edges incidents to the node.

$$C_d(i) = \deg(i) \qquad (7)$$

### 2.3.8. *Cluster coefficient:*

Cluster Coefficient is gained based on the following formula which A is the total number of edges between the nearest neighbors of vertex i and B is the maximum number of possible edges between the nearest neighbors of vertex i.[5]

$$CLUS_i = \frac{A}{B} \qquad (8)$$

The average of the Clustering Coefficients of all the vertices can be defined as Graph Cluster Coefficient.[3]

2.3.9. *Betweenness:*

Graph Betweenness centrality is perhaps one of the most prominent measures of centrality, it quantifies the number of times a vertex acts as a bridge along the shortest path between two other vertices.[3]

$$B_C(i) = \sum_{s \neq v_i \neq t} \frac{\sigma_{st}(v_i)}{\sigma_{st}} \quad (9)$$

In the above relation, $\sigma_{st}$ is the total number of shortest paths from vertex s to vertex t and $\sigma_{st(v_i)}$ is the number of those paths that pass through $v_i$.[3]

**2.4. *Hubs and motifs:***

We counted the Hubs and Motifs of both the normal and cancerous tissues. Highly connected nodes are usually defined as hubs. At present, definition of hubs is still an unsolved issue in biological network analysis.[2]

Several studies, for example [10, 11, 12] have considered different measures in order to defining the Hub. In a major advance in 2004, [10] supposed that nodes with the degree greater than 5 were labeled as hubs. [11] reported nodes with degree greater than 8 were labeled as hubs. [12] considered a degree cutoff of 20 was used to define hub proteins. In [13] the nodes with degree more than 12 were selected as hubs. Finally, In [4] nodes with degree greater than 8 were labeled as hubs and in [2] nodes with degree more than 5 and 12 were as hubs. In this paper, we apply four cutoffs (degree >5, degree >8, degree >12, degree >20) to define the hubs.

Motifs represent patterns in complex networks which occur significantly more often than in randomized networks. Some of them are important in order to specify functions in biological networks. Motif determination gives much of information about the properties and the characteristics of a network. One of the standard methods for comparing the PPI networks is based on the frequency of network motifs.[4]

**3. Experimental Results**

Since in biological networks, topological properties of nodes have important role to understanding the biological mechanisms, this paper provides a systematic analysis method of PPIs in both normal and cancerous tissue.

As mentioned earlier, the used dataset contains bone, breast, colon, kidney and liver PPI data of normal and cancerous tissues. For this purpose, such extracting and constructing of valid interactions was conducted using codes based on Python programming language. JetBrains PyCharm Community version was used for Python coding.

R 3.2.2 version has been used for all the network construction purposes, and RStudio for R Coding. Also, the iGraph package was used to determine the network parameters of each network. The protein interaction networks of normal and cancerous tissue for Breast, using network construction, has been depicted in Fig. 1.

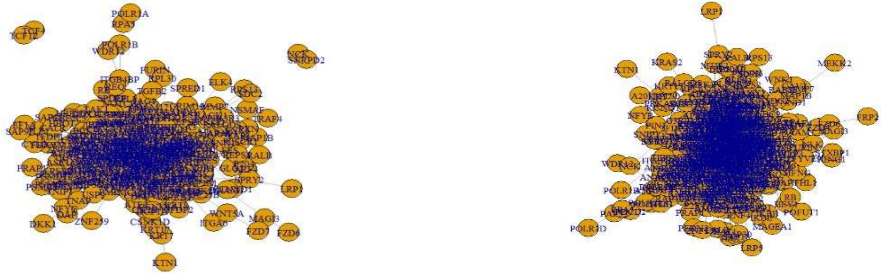

Fig. 1.  PPIs of normal and cancer tissue for Breast.

Specifications of data sets which are used in this study are demonstrated in Table 1 and Table 2. As it can be seen in Table 2, Number of edges in cancerous PINs decreases in comparison with normal tissues. This can be used as a measure for identification of cancerous tissues from normal tissues.

Table 1.  Vertices differences in Normal and Cancerous tissues

|  |  | Vertices | |
| --- | --- | --- | --- |
|  |  | Normal | Cancer |
| **Tissues** | Bone | 192 | 351 |
|  | Breast | 331 | 541 |
|  | Colon | 305 | 551 |
|  | Kidney | 315 | 491 |
|  | Liver | 302 | 631 |

Table 2.  Edges differences in Normal and Cancerous tissues

|  |  | Edges | |
| --- | --- | --- | --- |
|  |  | Normal | Cancer |
| **Tissues** | Bone | 619 | 1783 |
|  | Breast | 1696 | 3581 |
|  | Colon | 1492 | 4020 |
|  | Kidney | 1349 | 3140 |
|  | Liver | 1228 | 4478 |

We have calculated ten network parameters of closeness, graph diameter, graph radius, index of aggregation, connectivity, Number of Edges divided by the Number of Vertices, subgraph centrality, degree, cluster coefficient, betweenness to see how they have been changed in a normal and cancerous PPIs.

As the results in Table 3 show, Number of edges divided by the Number of vertices, Graph Degree Mean, Subgraph centrality, and Betweenness of the graph related parameters have significant differences in the same normal and cancerous tissues. Other parameters changes in different tissues of normal and cancerous. Therefore, some of the tissues have higher value in the normal cell while in other ones it's high for cancerous cell.

Table 3. Network parameters calculated for both the Normal and Cancerous tissues

|  |  |  | Tissues | | | | |
|---|---|---|---|---|---|---|---|
|  |  | Type | Bone | Breast | Colon | Kidney | Liver |
| **Parameters** | Closeness Centrality | Normal | 0.3222 | 0.3374 | 0.3384 | 0.3307 | 0.3181 |
|  |  | Cancer | 0.3335 | 0.3442 | 0.3448 | 0.3405 | 0.3379 |
|  | Graph Diameter | Normal | 7 | 9 | 8 | 7 | 8 |
|  |  | Cancer | 7 | 8 | 7 | 8 | 9 |
|  | Graph Radius | Normal | 4 | 1 | 4 | 1 | 4 |
|  |  | Cancer | 4 | 4 | 4 | 5 | 5 |
|  | Index of Aggregation | Normal | 1 | 0.9879 | 1 | 0.9841 | 1 |
|  |  | Cancer | 1 | 1 | 1 | 1 | 1 |
|  | Connectivity | Normal | 0.0337 | 0.0310 | 0.0321 | 0.0272 | 0.0270 |
|  |  | Cancer | 0.0290 | 0.0263 | 0.0265 | 0.0261 | 0.0225 |
|  | Number of edges by divided by the number of vertices | Normal | 0.1867 | 0.1614 | 0.1661 | 0.1857 | 0.1751 |
|  |  | Cancer | 0.1819 | 0.1888 | 0.1960 | 0.1653 | 0.1711 |
|  | Graph degree mean | Normal | 3.2239 | 5.17 | 4.89 | 4.33 | 4.06 |
|  |  | Cancer | 5.07 | 7.11 | 7.29 | 6.39 | 7.09 |
|  | Cluster Coefficient | Normal | 6.44 | 10.24 | 9.78 | 8.56 | 8.13 |
|  |  | Cancer | 10.15 | 14.23 | 14.59 | 12.79 | 14.19 |
|  | Betweenness | Normal | 211.4271 | 331.8066 | 310.8000 | 320.9112 | 337.6093 |
|  |  | Cancer | 364.3504 | 533.5767 | 542.1053 | 494.2281 | 640.9382 |
|  | Subgraph Centrality | Normal | $16*10^3$ | $367*10^6$ | $31*10^6$ | $10^6$ | $10^6$ |
|  |  | Cancer | $2*10^9$ | $17618*10^9$ | $82865*10^9$ | $638*10^9$ | $655589*10^9$ |

The other studied parameters within the PPIs of the normal and cancerous tissues were the hubs and motif. We have considered four different type of hub definitions. According to definitions, a hub is a node with more than five, eight, twelve, or twenty interactions. The number of hubs in the different tissues of normal and cancerous PINs are shown in Table 4.

Table 4. Hubs in the Normal and Cancerous Tissues

|  |  | Tissues | | | | |
|---|---|---|---|---|---|---|
|  | Type | Bone | Breast | Colon | Kidney | Liver |
| **Degree ≥** 5 | Normal | 71 | 188 | 181 | 159 | 151 |
| | Cancer | 204 | 365 | 372 | 324 | 404 |
| 8 | Normal | 40 | 133 | 117 | 108 | 91 |
| | Cancer | 150 | 291 | 302 | 245 | 310 |
| 12 | Normal | 26 | 79 | 75 | 60 | 58 |
| | Cancer | 83 | 209 | 216 | 156 | 212 |
| 20 | Normal | 12 | 40 | 36 | 26 | 21 |
| | Cancer | 43 | 108 | 111 | 80 | 115 |

As it can be seen in Table 4, the number of hubs in cancerous tissues are higher than their corresponding normal tissues. This results indicate that the cancer PINs are denser in hubs against their corresponding normal PINs.

We have also studied the motif differences of normal and cancerous tissues. We have considered with subgraph size 3. For this aim, we have used FANMOD to determine motif properties in normal and cancerous tissues PPIs. The motif ID and adjacency matrix is shown in Fig. 2.

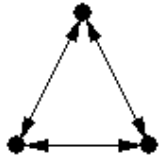

Fig. 2. Adjacency matrix of motif with ID 238.

The results in Table 5 reveal motif frequency differences in normal and cancerous tissues. This parameter changes in different tissues of normal and cancerous. As it seen in Table 5 motif with size 3 can not be used as a symptom for different of normal and cancerous tissues.

Table 5. Motifs frequency with size 3

|  | Tissues | | | | |
|---|---|---|---|---|---|
| Type | Bone | Breast | Colon | Kidney | Liver |
| Normal | 8.4783% | 10.304% | 9.6895% | 8.5608% | 9.768% |
| Cancer | 10.581% | 10.948% | 11.093% | 10.2% | 10.268% |

## 4. Discussion

Understanding the differences between normal and cancerous tissues is one of the critical problems in bioinformatics. We have studied the differences in graph related parameters of PPIs of both normal and cancerous tissues from different parts of the body including bone, breast, colon, kidney, and liver.

As it can be inferred from the results of this study, cancerous tissues of the five studied samples are denser in the interaction networks. That means their PPIs contains more edges in comparison with the normal networks of the same tissue.

The experimental results demonstrate that Number of edges divided by the number of vertices, Graph Degree Mean, Subgraph Centrality, Betweenness, and Hubs have higher values in the cancer cells and cancerous PPIs are denser more than of normal PPIs.

Therefore, these parameters are significantly different in the normal and cancerous PPIs and can be used as a measure to distinguish between normal and cancerous networks.

**Acknowledgments**

The authors like to acknowledge K. M. Taufiqur Rahman for their help and assistance.